\documentclass{mn2e}
\usepackage{graphicx}
\def\go{
\mathrel{\raise.3ex\hbox{$>$}\mkern-14mu\lower0.6ex\hbox{$\sim$}}
}
\def\lo{
\mathrel{\raise.3ex\hbox{$<$}\mkern-14mu\lower0.6ex\hbox{$\sim$}}
}
\def\simeq{
\mathrel{\raise.3ex\hbox{$\sim$}\mkern-14mu\lower0.4ex\hbox{$-$}}
}

\def\etal{{\it et al.\ }}

\def\etal{{\it et al.\ }}

\def\be{\begin{equation}}
\def\ee{\end{equation}}
\def\bea{\begin{eqnarray}}
\def\eea{\end{eqnarray}}

\def\etal{{\sl et al.\ }}

\def\hw2{{\hat W}^2}
\def\go{\mathrel{\raise.3ex\hbox{$>$}\mkern-14mu
             \lower0.6ex\hbox{$\sim$}}}
\def\lo{\mathrel{\raise.3ex\hbox{$<$}\mkern-14mu
             \lower0.6ex\hbox{$\sim$}}}
\def\ltorder{\mathrel{\raise.3ex\hbox{$<$}\mkern-14mu
             \lower0.6ex\hbox{$\sim$}}}
\def\gtorder{\mathrel{\raise.3ex\hbox{$>$}\mkern-14mu
             \lower0.6ex\hbox{$\sim$}}}

\def\eps2{{\epsilon^2}}

\def\la{\mathrel{\hbox{\rlap{\hbox{\lower4pt\hbox{$\sim$}}}{\raise2pt\hbox{$
>$}}}}}

\begin{document}

\title[Serendipitous AGN in the {\it XMM-Newton} fields of Markarian~205 and QSO~0130$-$403]
{Serendipitous AGN in the {\it XMM-Newton} fields of Markarian~205 and QSO~0130$-$403}
\author[K.L. Page \etal]{K.L. Page, M.J.L. Turner, J.N. Reeves,
P.T. O'Brien, S. Sembay\\
X-Ray Astronomy Group, Department of Physics \& Astronomy,  
Leicester, LE1 7RH, UK}

\date{Received ** *** 2002 / Accepted ** *** 2002}

\label{firstpage}

\maketitle

\begin{abstract}
The X-ray spectra of serendipitously observed AGN in the {\it XMM-Newton} fields of Mrk~205 and
QSO~0130$-$403 are analysed. The sample consists of 23 objects, none
of which is detected at radio frequencies, with a median X-ray luminosity of $\sim$4$\times
10^{44}$~erg~s$^{-1}$ and 
redshifts ranging from $\sim$0.1 to just over 3. The mean photon index
was found to be 1.89~$\pm$~0.04. In contrast to past {\it ASCA} and 
{\it ROSAT} observations of high-redshift 
radio-loud quasars, we find little evidence for excess intrinsic
absorption in these radio-quiet objects, with only three sources
requiring a column density in excess of the Galactic value.  
Comparing the measured spectral indices over the redshift
range, we also find there is no X-ray spectral evolution of QSOs with
time, up to redshift 3. Within the sample there is no evidence
for evolution of the optical to X-ray spectral index, $\alpha_{ox}$,
with redshift, the mean value being $-$1.66~$\pm$~0.04. However, upon
comparing the values from the Bright Quasar Survey at low redshift (z~$<$~0.5) and
high redshift QSOs detected by Chandra (z~$>$~4), 
a slight steepening of $\alpha_{ox}$ is noted for
the more distant objects. In most of the sources there is no significant
requirement for a soft excess, although a weak thermal component
($\leq$10~per~cent of L$_{X}$)
cannot be excluded. There is an
indication of spectral flattening (by $\Delta\Gamma=0.2$) 
at higher energies ($>3$ keV, QSO rest frame) for the sample as
a whole. This is consistent with the presence of a Compton reflection 
component in these radio-quiet AGN, with the scattering 
medium (such as an accretion disc or molecular torus) occupying a 
solid angle of $2\pi$ steradians to the X-ray source.

\end{abstract}


\section{Introduction}
\label{sec:intro}

The majority of luminous AGN (Active Galactic Nuclei) are found to be 
radio-quiet (e.g., Kukula \etal 1998); that is, the
radio (5~GHz) to optical ({\it B}-band) flux ratio is $\leq$~10
(Wilkes \& Elvis 1987; Kellerman \etal 1989). 
The radio emission of radio-quiet QSOs is
not non-existent, but is about 100 times less than that of the
radio-loud quasars. In general, for a given optical luminosity, the
average X-ray emission from radio-quiet AGN is about three times lower
than that from radio-loud objects (Zamorani \etal 1981; Worrall \etal
1987). For this
reason, they are more difficult to detect at higher redshifts, so the
majority of distant AGN studied so far in the X-ray band are
radio-loud. One method of increasing the number of high-z radio-quiet
objects observed is to make use of the field sources in deep observations taken
with a sensitive instrument, such as {\it XMM-Newton}. In this paper we
present a study of the background field sources from two long {\it
XMM-Newton} observations; for both fields optical identifications exist, so
that the redshift and optical classification of each object are known.

Although unified models exist, claiming that most AGN subclasses can be 
understood simply through orientation
effects (see Antonucci 1993 and references therein), the difference between 
radio-quiet and radio-loud QSOs
(RQQs and RLQs) has yet to be completely explained. 

It was once thought that RQQs were only found in spiral galaxies,
while the RLQs inhabited ellipticals. However, advances in observing
capability now show that as many as 50~per~cent of the
RQQs discovered may occur in ellipticals (e.g., V\'{e}ron-Cetty \&
Woltjer 1990). RLQs, however, are almost exclusively found in bulge-dominated systems; that is, in ellipticals or early-type
spirals (Falomo, Kotilainen \& Treves 2001).

The {\it Einstein} IPC (Imaging Proportional Counter) data showed that
radio-quiet and radio-loud
objects have different power law slopes over the energy range 
0.4--4~keV, with the RQQs being
steeper (e.g., Wilkes \& Elvis 1987). Follow-up with {\it ROSAT}
(Brinkmann, Yuan \& Siebert 1997; Yuan \etal 1998) confirmed these 
findings for the soft
X-ray band and {\it Ginga} (Williams \etal 1992), {\it EXOSAT}
(Lawson \etal 1992) and {\it
ASCA} (Reeves \etal 1997; Reeves \& Turner 2000 -- RT00 hereafter) 
for the harder X-ray energies.
This observed difference in slope ($\Delta\Gamma$~$\sim$~0.3 over
2--10~keV -- Lawson \etal 1992; $\Delta\Gamma$~$\sim$~0.5 in the
0.2--3.5~keV band -- Wilkes \& Elvis 1987) may be due 
to an extra, hard X-ray component linked to the radio emission of
RLQs; i.e., connected with the radio-jets. 

Neither {\it Einstein} (Canizares \& White 1989; Wilkes \& Elvis
1987) nor {\it ROSAT} (Yuan \etal 1998) found any indication for
excess absorption in radio-quiet QSOs. Yuan \etal (1998) also found
that the mean spectral slope of nearby RQQs is consistent with that of
more distant (z~$>$~2.5) QSOs, implying that X-ray spectral evolution
may be unimportant in the radio-quiet objects.

\begin{figure}
\begin{center}
\includegraphics[clip,width=6.0cm,angle=-90]{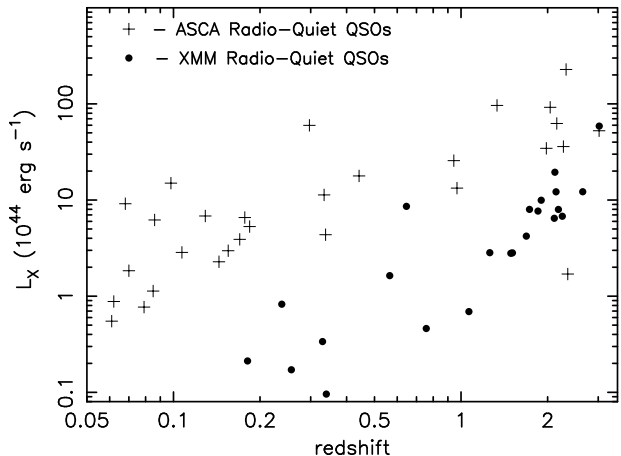}
\caption{The luminosities of radio-quiet objects are plotted against
redshift. Measurements marked as crosses are taken from Reeves \&
Turner (2000); circles are from this paper.}
\label{lumdif}
\end{center}
\end{figure}

Because of the effective area of its mirrors, {\it XMM-Newton} can
obtain spectra of fainter objects for a given redshift than could any previous
X-ray instrument. This allows us to extend previous surveys of QSOs to
include those objects with lower luminosity and to observe
a given luminosity at a wider range of redshifts. In this respect, the
luminosities of the objects presented in this paper (median value of
$\sim$~4~$\times~10^{44}$ erg~s$^{-1}$) are closer to those of the AGN
which make up the cosmic X-ray background. 
 Fig.~\ref{lumdif} shows the luminosity distribution of 
radio-quiet objects surveyed by {\it XMM} in this paper compared, with
those surveyed by {\it ASCA} in RT00. 

In Section~\ref{sec:xmmobs} the data reduction is discussed, with the
spectral fitting to the individual objects outlined in
Section~\ref{sec:specanal}. The highest-flux objects are analysed in
\ref{sec:high}, while the work on Mrk~205 itself was
presented in a separate paper (Reeves \etal 2001). 
Finally, the results are discussed in Section~\ref{sec:disc}.

\section{XMM-Newton Observations}
\label{sec:xmmobs}

The objects in this paper were serendipitously located in the fields of
two {\it XMM} pointed observations: Mrk 205 and QSO~0130$-$403. The 
Mrk~205 data are from a calibration observation, in May
2000 (OBS ID 0124110101). There were three separate exposures with 
both the MOS and PN
cameras (Turner \etal 2001; Str\"{u}der \etal 2001), each of which 
lasted for $\sim$17~ksec, and these were co-added before spectral analysis was
performed. QSO~0130$-$403 was observed only once, for $\sim$37~ksec, 
in June 2001 (OBS ID0112630201).

The optical
identifications and redshifts for the Mrk~205 field are taken from
Barcons \etal (2002), henceforth XB02, which also quotes X-ray colours
and fluxes. For the QSO~0130$-$403
field, the IDs were taken from the V\'{e}ron-V\'{e}ron 2001
catalogue. Hence, the Mrk~205 objects were selected on the basis of
their X-ray flux ($>$2~$\times~10^{14}$~erg~cm$^{-2}$~s$^{-1}$ in the
0.5--4.5~keV band; XB02), with all objects down to this flux level and
not in a CCD chip-gap included. The objects analysed in the second
field were those which had existing optical identifications; all the
optically-selected QSOs within the {\it XMM} field of view were
detected and included in the sample.

In general, one might expect that X-ray-selected targets would show
harder spectra than those found through optical observations, simply
because soft X-ray sources are not as easy to detect, due to
Galactic absorption. Comparing both the slopes over 2--10 (rest
frame) and 0.2--10~(observed frame)~keV for the objects in the two
separate fields,
a mean difference of $\Delta\Gamma$~$<$~0.1 is found. The actual range of
photon indices is also very similar for both of the fields, with the vast
majority ($\sim$80~per cent) of the spectral slopes lying between
$\Gamma$~$\sim$~1.6--2.4. The similarity in the distributions of
indices implies that there is no obvious problem in combining these two fields.

The data obtained allow individual spectral fitting for the QSOs in
these fields. We present these spectra and investigate the sample as a
whole, to determine
whether there is any evidence for evolution, Compton reflection or
excess absorption in the rest frames of the objects. A wide range of
redshifts, from z~=~0.181 to 3.023, is covered.

The data were reduced with the {\it XMM} SAS (Science Analysis
Software) version 5.2, using {\sc epchain} and {\sc emchain} to generate valid
photon lists; further filtering was then performed using {\sc
xmmselect}. Both single and double pixel events (patterns 0--4) were
used when extracting the PN spectra, while patterns 0--12 were chosen
for the MOS exposures. Each spectrum was extracted using a circular 
region of 30
arcsec radius, with the same circle then offset from the source to
obtain an estimate of the background. 

Since the sources were not all
close to the centre of the CCDs, {\sc arfgen} was run within the SAS,
to correct for vignetting effects and the encircled energy
fraction. Subsequently, CCD redistribution matrices (i.e., {\it .rmf} 
not {\it .rsp}) were used. The energy
resolution of the PN chips changes as one moves out from the centre,
due to charge transfer effects; to take this into account, PN rmfs are
provided at intervals of 20 rows, with Y0 being used for rows 1-20
(furthest from the centre) up to Y9 for the central objects. For
example, epn$\_$ff20$\_$sdY9$\_$qe.rmf was used with {\sc arfgen} for
a central source (Y9), extracted for patterns 0-4 (sd~=~singles and doubles) in
a full-frame (ff) image. Note that, for pure redistribution matrices,
the type of filter (thin, medium or thick) is not included; this
becomes part of the ancillary response function (arf) generated.
For MOS, the relevant responses were m1$\_$r5$\_$all$\_$15.rmf and
m2$\_$r5$\_$all$\_$15.rmf ({\it all} meaning the
spectra were extracted for patterns 0--12). See {\bf http://xmm.vilspa.esa.es/ccf/epic} for more details.

After using the {\sc ftool} command {\sc grppha}, to provide a
minimum of 20 counts per bin, the {\sc Xspec} v11.0.1 software package
was used to analyse the background-subtracted spectra. 
The value of the Galactic absorption was obtained using the
{\sc ftool} {\sc n$_H$}, which uses data from Dickey \& Lockman
(1990). N$_H$ was found to be 3~$\times~10^{20}$~cm$^{-2}$ for the 
Mrk~205 field
and 1.9~$\times~10^{20}$~cm$^{-2}$ for the QSO~0130$-$403 field. H$_0$
was taken to be 50~km~s$^{-1}$~Mpc$^{-1}$ and q$_0$ to be 0.5.
Unless stated otherwise, the errors given are at the 1~$\sigma$ level.

Of the 23 AGN analysed in this paper, only two (XMMU J013257.8-401030 and
XMMU J013333.8-395554) have actual (upper limit) radio observations, of
0.4 mJy at 5~GHz (Miller, Peacock \& Mead 1990). None of the
others are detected in the NVSS (NRAO VLA Sky Survey), down to the
survey completeness flux limit of about 2.5 mJy. Using this flux limit (and the two actual upper limits), we find
that six of the AGN are definitely radio-quiet. Although the remaining seventeen
objects could be radio-loud, it is found that (again using the NVSS
survey limit) most would only be slightly so, with twelve having R~$<$~60.

Statistically, one would expect $\sim$20~per~cent of the BLAGN (with X-ray fluxes above
10$^{-14}$~erg~cm$^{-2}$~s$^{-1}$) in an XMM field to be radio loud
(based on studies of sources within the {\it XMM} observation of the
Lockman Hole; private communication, Lehmann 2002); this corresponds to $\sim$4 objects in a
sample of this size. However, as our survey does not go as deep as
10$^{-14}$~erg~cm$^{-2}$~s$^{-1}$, it should be expected that this is
an upper limit to the number of radio-loud sources.

\section{Spectral Analysis}
\label{sec:specanal}

As is conventional, the spectrum for each source was first modelled using
a simple power law, together with neutral hydrogen absorption fixed at
the Galactic value. The majority of sources were seen
with both the
MOS and PN, although the occasional source would be positioned in
one of the chip gaps, or over a bad column. Where both MOS and PN CCDs
could be used, joint fits were performed. For sixteen of the QSOs, the single
power law model produced a good fit, with
$\chi^{2}_{\nu}$~$\sim$~0.8--1.2. The remaining seven objects showed
slightly worse statistics, but the fits were still acceptable.

Table~\ref{sources} lists the redshift and optical magnitude (from XB02 for the
Mrk~205 field and NED -- the NASA Extragalactic Database -- for the
objects around QSO~0130$-$403), count-rate, flux, luminosity and 2--10~keV (QSO rest frame)
spectral slope (all derived from our data) for
each of the 23 sources. All but three of the objects in this sample
are Broad Line AGN (BLAGN). XMMU~J122120.5+751616 and 
XMMU~J122206.4+752613 are classified as Narrow Emission Line AGN (NLAGN)
by XB02, with XMMU~J122258.1+751934 tentatively classified as
the same. It should be noted that the optical spectrum for this last
object is not of high quality and there is, therefore, some
uncertainty in the given redshift of 0.257. It can be seen from the
column giving the luminosities that all except two of the
objects have a 2--10~keV luminosity of $>$10$^{43}$~erg~s$^{-1}$; they
should, therefore, be classified as QSOs, rather than any of the lower-luminosity objects which come
under the title of AGN.

\begin{table*}
\begin{center}
\caption{QSOs with known redshifts in the fields of Mrk~205 and
QSO~0130$-$403. The luminosity was calculated for q$_0$~=~0.5 and
H$_0$~=~50~km~s$^{-1}$~Mpc$^{-1}$; the absorption-corrected value is given
for the QSO rest-frame 2--10~keV band. The optical magnitudes given
are g' for the Mrk~205 field objects and V-band for QSO~0130-403. 
Alternative names for the QSOs in
the QSO~0130$-$403 field are given at the foot of the table. VV01 indicates
the V{\'e}ron-V{\'e}ron catalogue (V{\'e}ron-Cetty \& V{\'e}ron 2001)} 
\label{sources}
\begin{tabular}{p{3.8truecm}p{0.8truecm}p{2.0truecm}p{2.2truecm}p{2.0truecm}p{2.0truecm}p{3.0truecm}}
Source ID & z &0.2--10 keV  &2--10 keV Flux& 2--10 keV & Optical & 2--10 keV \\
 & & Count Rate$^{1}$ &(10$^{-14}$
 & Luminosity & Magnitude & rest frame $\Gamma$ \\
&  & (10$^{-3}$ s$^{-1}$) &erg cm$^{-2}$ s$^{-1}$) & ($10^{44}$ erg s$^{-1}$) &
\\
&\\
XMMU J121819.4+751919& 2.649 & 7.39 $\pm$
1.10 & 1.49  &38.20  & 19.79
& 2.05~$\pm$~0.41\\
XMMU J122048.4+751804 & 1.687 &  13.01 $\pm$
1.19 & 3.13 & 4.25 & 18.97
& 1.95~$\pm$~0.36\\
XMMU J122051.7+752820 & 0.181 &2.95 $\pm$
1.11& 1.32  & 0.022 & 22.05 &1.64~$\pm$~1.37\\
XMMU J122052.0+750529 & 0.646& 65.75 $\pm$
1.91 & 41.97  & 8.63 & 18.62
& 1.68~$\pm$~0.10\\
XMMU J122111.2+751117 & 1.259 &17.83 $\pm$
1.17 & 2.56  & 2.87 & 19.54
& 2.13~$\pm$~0.30\\
XMMU J122120.5+751616 & 0.340& 16.28 $\pm$
1.23 & 1.60  & 0.096 & 20.90
& 1.82~$\pm$~0.44 \\
XMMU J122135.5+750914 & 0.330 &13.78 $\pm$
1.15 & 5.85 & 0.29 & 20.38
 & 1.26~$\pm$~0.26\\
XMMU J122206.4+752613 & 0.238 & 115.2 $\pm$
1.89  & 36.85 & 0.91 & 20.03
& 1.77~$\pm$~0.77\\
XMMU J122242.6+751434 & 1.065 & 6.05 $\pm$
1.14& 0.35  & 0.28 & 21.82
& 2.61~$\pm$~0.36\\
XMMU J122258.1+751934 & 0.257 &12.86 $\pm$
1.20 & 7.07  & 0.20 & 22.66
& 1.74~$\pm$~0.29\\
XMMU J122318.5+751504 & 1.509 & 7.72 $\pm$
1.13 & 1.70  & 3.00 & 21.15
& 1.43~$\pm$~0.29\\
XMMU J122344.7+751922 & 0.757 & 9.87 $\pm$
1.26 & 1.01  &0.46 & 20.69
& 1.77~$\pm$~0.59\\
XMMU J122351.0+752227 & 0.565 & 33.81 $\pm$
1.71 & 10.17 &1.63 & 19.93
& 1.72~$\pm$~0.17\\
XMMU J122445.5+752224 & 1.852 &9.68 $\pm$
1.40 & 3.80 & 7.67 & 20.16
& 2.17~$\pm$~0.36\\
XMMU J013302.0$-$400628$^{a}$ & 3.023 & 74.44 $\pm$
1.89 & 7.63 & 82.80 & 17.4
& 2.10~$\pm$~0.07\\ 
XMMU J013257.8$-$401030$^{b}$ & 2.180 & 17.96 $\pm$
1.42 & 2.74 & 7.46 & 19.2
& 1.67~$\pm$~0.23\\
XMMU J013314.7$-$401212$^{c}$ & 1.49 & 4.82 $\pm$
0.43 & 1.09 & 2.82 & 19.0 & 2.27~$\pm$~0.58\\
XMMU J013205.4$-$400047$^{d}$ & 2.12 &  25.82 $\pm$
1.46 & 4.57 &18.60 & 20.16
& 2.05~$\pm$~0.16\\
XMMU J013348.3$-$400233$^{e}$ & 2.11 & 2.49 $\pm$
0.37& 3.68 &5.87 & 20.3
& 3.48~$\pm$~1.14\\
XMMU J013233.2$-$395445$^{f}$ & 1.73 & 3.14 $\pm$
0.37 & 5.37 &7.49 & 20.3
& 1.77~$\pm$~0.38\\
XMMU J013333.8$-$395554$^{g}$ & 1.90 &3.98 $\pm$
0.41 & 3.52 &9.41 & 19.8
& 2.21~$\pm$~0.32\\
XMMU J013408.6$-$400547$^{h}$ & 2.14 &17.06 $\pm$
0.13& 1.90 & 12.30 & 19.56
& 2.33~$\pm$~0.25\\
XMMU J013320.4$-$402021$^{i}$ & 2.25 &4.37 $\pm$
1.06 & 3.87  & 6.98 & 20.5
& 1.37~$\pm$~0.61\\

\end{tabular}
\end{center}
$^{1}$~PN count rate, unless MOS data only available.\\
$^{a}$~QSO~0130$-$403; $^{b}$~[VV01]~J013257.5-401028; $^{c}$~[VV01]~J013315.2-401203; $^{d}$~[VV01]~013205.4-400048; $^{e}$~[VV01]~J013348.2-400235; $^{f}$~[VV01]~J013233.2-395445; $^{g}$~[VV01]~013334.1-395546;
$^{h}$~[VV01]~013408.5-400547; $^{i}$~[VV01]~J013320.4-402022
\end{table*}

\begin{figure}
\begin{center}
\includegraphics[clip,width=6.0cm,angle=-90]{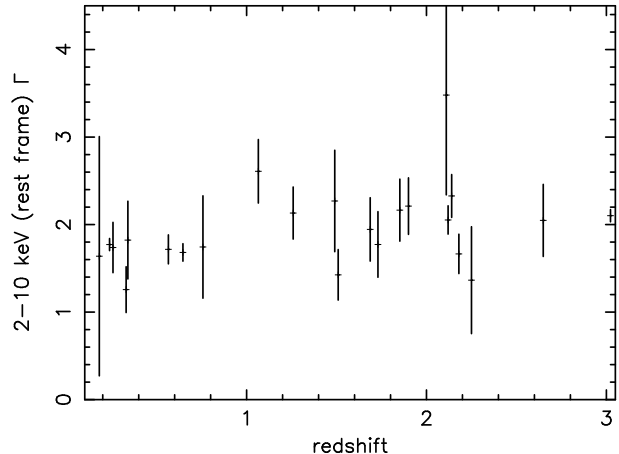}
\caption{Plot of 2--10~keV $\Gamma$ against redshift for each of the 
23 sources. No clear trend between $\Gamma$ and z is found 
for the sample.}
\label{evol}
\end{center}
\end{figure}

Over the 2--10~keV (rest-frame) band, the weighted mean of the 
photon index for all 23 objects was found to be
$\Gamma$~=~1.89~$\pm$~0.04. The dispersion of the slopes from the 
mean, taking into account the measurement errors, is $\sigma
\sim$~0.21 (see Fig~\ref{evol}). Using the more rigorous Maximum
Likelihood method (Maccacaro \etal 1988), the dispersion was calculated to be 0.29~$\pm$~0.04.
The mean of the 2--10~keV photon index for the RLQs in RT00 is given
as 1.66~$\pm$~0.04, while the RQQs had a mean $\Gamma$ of
1.89~$\pm$~0.05. It can be seen that the {\it
XMM}-observed radio quiet objects in this paper have a mean slope
which is in excellent agreement with the higher flux objects 
measured by {\it ASCA} and that the
radio-loud {\it ASCA} objects have a flatter index ($\Delta\Gamma$~$\sim$~0.2),
corroborating the known difference between radio-loud and radio-quiet 
QSOs found in previous surveys.

\subsection{Soft Excess and Absorbing Column Density}
\label{sec:specanal_nh}

For each source, a low-energy blackbody component was added, 
to see if there was
a measurable soft excess; however, it was found that, statistically,
only one of the
sources required this addition: XMMU~J122135.5+750914 was better
fitted by the inclusion of a blackbody component with
kT~=~0.152~$\pm$~0.031~keV. The luminosity of the soft
excess component was $\sim$18~per~cent that of the power law.

Excess absorption in the QSO rest
frame was also investigated, using the {\it zwabs} model in {\sc
Xspec} to include a component in addition to the Galactic column
density. Only three of the sources in the joint fields showed any
improvement in the reduced {$\chi^{2}$}, one of which
(XMMU~J122344.7+751922) then gained an
extremely steep ({$\Gamma~\sim$~3.5}) spectral slope, for an excess
absorption of (2.27~$\pm$~1.06)~$\times~10^{21}$~cm$^{-2}$. The other two
objects, both classed as NLAGN, were XMMU~J122206.4+752613, 
with an improvement in
{$\chi^{2}$} of 40, for one degree of freedom, for an additional column
density of (6.0~$\pm$~1.0)~$\times~10^{20}$~cm$^{-2}$ and
XMMU~J122258.1+751934, showing a decrease in {$\chi^{2}$} of six for one
degree of freedom, for N$_{H}$~=~(1.28~$\pm$~0.64)~$\times~10^{21}$~cm$^{-2}$.
All three of these objects are of low luminosity.

Although the higher redshift objects appeared
to show larger upper limits for the absorbing column density, it
should be noted that this is
likely to be related to the shift in energy bandpass. For an object
with z~=~3, the 0.2--10~keV observed band corresponds to 0.8--40~keV
in its rest frame. It is, therefore, obvious that there will be
increased difficulty in measuring moderately small absorption 
related to the softest X-ray photons and, so, a higher upper limit will
be found compared to a source of the same flux at a lower redshift.

We also tried fixing the photon index to a typical value of $\Gamma=1.9$
(e.g., Nandra \& Pounds 1994), to see whether the 
apparent dispersion in $\Gamma$ could be explained by excess
absorption in the QSO rest frame; i.e., to see if any of the flat spectral
slopes were due to the objects being obscured. The model
comprised a direct, absorbed power law, together with a scattered,
unabsorbed component (both with $\Gamma$~=~1.9). However, for ten of the 
objects in the sample,
this produced a worse fit than that with no obscuration component 
and only four sources were significantly better-fitted
using this model -- the two NLAGN which required the {\it
zwabs} column density, together with XMMU~J122052.0+750529 and 
XMMU~J122135.5+750914. The
obscuring columns found for the NLAGN were (2.40~$\pm$~0.57) and
(1.75~$\pm$~1.57)~$\times~10^{21}$~cm$^{-2}$, while the BLAGN
objects required (7.73~$\pm$~4.88) and (17.88~$\pm$~11.85)~$\times~10^{22}$~cm$^{-2}$ respectively. The
remaining nine objects can be equally well fitted by either a simple
power law, or by including the obscured component. It should be
noted that the objects with particularly flat spectral indices are all
included in this group. 

Although, statistically, these nine spectra can be fitted by the
inclusion of an obscuring column density, the values for N$_{H}$ are
not well constrained and are entirely consistent with zero. 
This implies that the 
more complicated model is not required and
that these QSO nuclei are not heavily absorbed, consistent with AGN 
unification schemes where one should observe a direct line to the nucleus 
in these type-1 AGN. To summarise, there appears to be an  
intrinsic dispersion in the primary X-ray power
law slopes of these background AGN, which cannot be solely caused by 
line of sight absorption. 

In order to try and obtain better statistics for the fits, it was 
decided to co-add
spectra over three redshift ranges -- z~=~0--1, 1--2 and $>2$ -- and
look at the subsequent continuum shapes. For each grouping, an average
background spectrum was produced and the mean redshift
calculated. Although co-adding spectra will tend to blur sharp
features, such as iron emission lines, this should not be a problem
when simply looking at the continuum as a whole.

Only the resulting spectrum for the low-redshift group (mean z~=~0.4) required excess
absorption, with an additional column density of 
(2.9~$\pm$~0.9)~$\times~10^{20}$~cm$^{-2}$ for $\Gamma$~=~1.83~$\pm$~0.03. 
Such a column would be undetectable at higher redshifts, as it would be 
shifted out of the {\it XMM-Newton} bandpass. Even with the better
statistics derived from the co-adding of the spectra, there was still
no requirement for a blackbody component for either the low- or medium-redshift group, with the luminosity of the
soft component being $<$6~per~cent of the
total X-ray luminosity. For the highest-redshift objects, however, a
blackbody does improve the fit significantly: a temperature of
kT~=~0.214~$\pm$~0.043 leads to a reduction in $\chi^{2}$ of 8, for 2
degrees of freedom ($>$90 per cent confidence, using the F-test), with
the luminosity of the BB component being almost 15~per~cent that of
the power law. However, as is discussed in
Section~\ref{sec:specanal_speccurv}, some of this spectral curvature
could be due to reflection. 

As an alternative method for identifying curvature due to a soft
excess, the lowest-redshift co-added spectrum was fitted with a broken
power law, with the break at 2~keV in the rest frame. $\Gamma$ was
measured to be 1.86~$\pm$~0.10 below the break and 1.82~$\pm$~0.05
above. Hence, there is little steepening of the spectrum at lower energies, the upper
limit being $\Delta\Gamma$$<$0.18; this is consistent with the finding above, that any soft
excess present in the low redshift objects is only weak.

\subsection{Spectral Evolution}
\label{sec:specanal_specevol}

Fig.~\ref{evol} shows the 2--10~keV power law indices versus
redshift. The dispersion in $\Gamma$ is clearly visible,
but there is no noticeable trend with z. A basic Spearman-Rank test to 
the broad-band photon index versus redshift data implied there was no correlation 
between the values.

Weighted linear regression was then performed, producing a best-fit
slope to the $\Gamma$-z graph of 0.14~$\pm$~0.03. However, five
measurements have rather small errors and so may be biasing the
result; these were, therefore, removed and the statistic recalculated.
This gave a value for the gradient of
0.07~$\pm$~0.12, so implying that there is little, or no, correlation between
the spectral slope and the redshift for the objects analysed. This
lack of correlation is also noted for the $\Gamma$-luminosity data, as
would be expected from the tight correlation between L$_{X}$ and redshift.

Values of the two-point optical to X-ray spectral index, 
$\alpha_{ox}$, were measured, where

\begin{displaymath}
\alpha_{ox} = \frac{log\frac{f_{\nu}(2 keV)}{f_{\nu}(2500
\AA)}}{log\frac{{\nu}(2 keV)}{{\nu}(2500 \AA)}}
\end{displaymath}

 and the mean calculated to be
$-$1.66~$\pm$~0.04. For this sample alone, no correlation was found
between $\alpha_{ox}$ and redshift. 
However, previous results from the Bright
Quasar Survey (BQS) gave a slightly flatter value of 
$\alpha_{ox}$~=~$-$1.56~$\pm$~0.02, for
unabsorbed QSOs at low redshift (with z~$<$~0.5), while the mean slope found by
Chandra for the more distant (z~$>$~4) objects is somewhat steeper,
at $-$1.78~$\pm$~0.03 (both values from Vignali \etal 2001). Comparing these
measurements from the BQS, Chandra and {\it XMM}, there is an indication that $\alpha_{ox}$ steepens slightly with
increasing redshift, with the extreme high redshift QSOs (at z~$>$~4)
being relatively fainter in the X-ray band, compared to low-z
objects. One should be cautious when comparing samples of
optically-selected (e.g., BQS) and X-ray-selected
objects, since selection biases may lead to complications. However, some
studies (e.g., Lamer, Brunner \& Staubert 1997) have shown that the
differences are not always significant. Likewise, as mentioned in
Section~\ref{sec:xmmobs}, the differently-selected fields
in this paper show a consistent range of spectral parameters.

\subsection{Spectral Curvature}
\label{sec:specanal_speccurv}

AGN, as a group, often show `reflection' features, due to
back-irradiation of the accretion disc by the hard X-ray flux (Pounds
\etal 1990; Nandra \& Pounds 1994), the strength of the features
depending on the solid angle subtended by the optically thick,
reflecting material. The
strongest of these features is often an emission line arising
from iron fluorescence, at $\sim$6.4--6.9~keV, depending on the
ionisation state of the material. An absorption edge, located at
7.1~keV (for neutral iron), should also be visible. The 
individual spectra we obtained show no obvious
features around this energy range, possibly because of limited
signal-to-noise. 
The `Compton reflection hump' (which flattens the high energy spectrum)
occurs due to the decreasing photo-electric absorption opacity and is
expected to peak around 20--40~keV in the rest frame of the object.  Because
of the high redshift of the objects in this sample, the observed {\it
XMM} energy band extends to up to 30--40~keV in the rest
frame and, so, the spectra might be expected to show evidence of the Compton hump.

Neutral reflection models were tried ({\it pexrav} in {\sc
Xspec}; Magdziarz \& Zdziarski 1995), fixing the exponential energy cut-off
at 100~keV and the disc inclination angle at 30$^{o}$. The strength
of the reflection component is given by R~=~$\Omega$/2$\pi$, where
$\Omega$ is the solid angle subtended by the reflector. Upon fitting 
the AGN individually, only one of the brightest objects,
QSO~0130$-$403, showed evidence for reflection, with $\chi^{2}$
decreasing by 16 for 1 degree of freedom, corresponding to a
probability of $>$99.99~per~cent via an F-test.

A simpler method, more appropriate
to the limited data quality, was to try and identify spectral curvature 
by measuring the photon index over two distinct energy bands 
(defined as 0.5--3~keV
and 3+~keV in the rest frame of the object) and hence determine the
difference between the `hard' and `soft'-band values.
The values of the slope in all energy bands
analysed are shown in Table~\ref{gammas}. At the foot of each column,
the mean value for the photon index in that energy 
range is given. 

If Compton reflection is present, then one would expect the higher-energy
spectral slope to be flatter, hence subtracting the soft-band $\Gamma$ 
from the harder one would lead to a negative value. Simulations were
performed, for an object at the mean redshift of our sample (z~=~1.4),
setting R~=~1 (corresponding to reflection from the full
2$\pi$ steradians of Compton thick matter); $\Gamma$ was then measured 
over the two energy bands defined above (0.5--3~keV and 3+~keV) and the difference in photon index
over the two bands was calculated to be $\Delta\Gamma$~$\sim$~$-$0.17.
The weighted mean for the difference in
the slopes for our actual spectra was found to be $-$0.15~$\pm$~0.06, which
corresponds to a reflection component, R, of 0.88~$\pm$~0.35. The
spectral shape for the sample as a whole is therefore consistent 
with reflection from approximately 2$\pi$ steradians of optically
thick matter, located out of the direct line of sight (such as an
accretion disc or molecular torus).

As when investigating the presence of a soft excess and additional
absorption, the coadded spectra over three redshift bands were analysed. The photon
indices were measured over 0.5--3~keV (rest frame) and 3+~keV energy bands, to
compare with the findings above. For the lowest-redshift objects, there
was no noticeable change in slope; however, flattenings at the higher-energies of
$\Delta\Gamma$~=~$-$0.25~$\pm$~0.18 and $-$0.30~$\pm$~0.09 were calculated for
the middle- and high-redshift
groups respectively. This
prompted the use of {\it pexrav}, which indicated the presence of a 
reflection component (detected at 98~per~cent significance), 
of R~$\la$~0.9 for the most distant objects; there was little
improvement in the fit for the medium-z group. This is consistent with the result obtained from finding the average change in slope between the soft and
hard energy bands for the individual spectra.

In Section~\ref{sec:specanal_nh}, the possibility of a soft excess was
discussed. Adding in a blackbody component to the model for the high-z
group gave an
acceptable fit, although, statistically, not as good as for
reflection. It also seems unlikely that there would be a more
clearly 
detectable soft excess at higher redshifts than the lower-z values, which are more sensitive to any excess at the lowest
energies. Nonetheless, a model comprising both a blackbody and a
reflection component still improved the fit, giving a
$\chi^{2}_{\nu}$~=~142/144, corresponding to almost 97~per~cent
confidence for the addition of the reflection component.
Thus, there does appear to be an indication of the Compton reflection
hump, in the most distant objects.

\begin{table*}
\begin{center}
\caption{Power law slopes over a range of energy bands for the sources
in the combined Mrk~205 and QSO~0130$-$403 fields.} 

\label{gammas}
\begin{tabular}{p{4.0truecm}p{0.7truecm}p{2.0truecm}p{2.0truecm}p{2.0truecm}p{1.5truecm}}
Source ID & z & 2--10 keV & 0.5--3 keV & 3+ keV & 0.2--10 keV \\
& &rest frame  $\Gamma$ & rest frame $\Gamma$ & rest frame $\Gamma$ & obs. frame $\Gamma$\\
&\\
XMMU J121819.4+751919 & 2.649 & 2.05 $\pm$ 0.41& 3.36 $\pm$ 0.72 & 2.44 $\pm$
0.45 & 2.29 $\pm$ 0.18 \\
XMMU J122048.4+751804 & 1.687 & 1.95 $\pm$ 0.36& 2.21 $\pm$ 0.38 & 1.54 $\pm$
0.32 & 1.66 $\pm$ 0.15 \\
XMMU J122051.7+752820 & 0.181 & 1.64 $\pm$ 1.37 & 2.37 $\pm$ 0.39 & 4.28 $\pm$
3.04 & 1.94 $\pm$ 0.25\\
XMMU J122052.0+750529 & 0.646& 1.68 $\pm$ 0.10 & 1.83 $\pm$ 0.07 & 1.74 $\pm$
0.15 & 1.77 $\pm$ 0.04  \\
XMMU J122111.2+751117 & 1.259 &2.13 $\pm$ 0.30 & 2.08 $\pm$ 0.16 & 1.30 $\pm$
0.33 & 2.18 $\pm$ 0.10 \\
XMMU J122120.5+751616 & 0.340& 1.82 $\pm$ 0.44 & 2.34 $\pm$ 0.18 & 1.93
$\pm$ 1.04 & 2.19 $\pm$ 0.12 \\
XMMU J122135.5+750914 & 0.330 &1.26 $\pm$ 0.26 & 1.91 $\pm$ 0.17 & 1.48 $\pm$
0.50 & 1.69 $\pm$ 0.10 \\
XMMU J122206.4+752613 & 0.238 & 1.77 $\pm$ 0.07 & 1.50 $\pm$ 0.04 & 1.72 $\pm$
0.13 & 1.56 $\pm$ 0.02 \\
XMMU J122242.6+751434 & 1.065 & 2.61 $\pm$ 0.36 & 2.19 $\pm$ 0.81 & -0.73 $\pm$
2.05 & 2.40 $\pm$ 0.52 \\
XMMU J122258.1+751934 & 0.257 &1.74 $\pm$ 0.29 & 1.46 $\pm$ 0.18 & 2.00 $\pm$
0.61 & 1.41 $\pm$ 0.09\\
XMMU J122318.5+751504 & 1.509 & 1.43 $\pm$ 0.29 & 2.41 $\pm$ 0.30 & 2.30 $\pm$
0.49 & 2.05 $\pm$ 0.15  \\
XMMU J122344.7+751922 & 0.757 & 1.74 $\pm$ 0.59 & 2.67 $\pm$ 0.17 & 2.24 $\pm$
0.98 & 2.63 $\pm$ 0.13 \\
XMMU J122351.0+752227 & 0.565 & 1.72 $\pm$ 0.17 & 1.94 $\pm$ 0.09 & 1.39 $\pm$
0.26 & 1.88 $\pm$ 0.05\\
XMMU J122445.5+752224 & 1.852 & 2.17 $\pm$ 0.36 & 1.26 $\pm$ 0.44 & 2.08 $\pm$
0.43 & 1.87 $\pm$ 0.16 \\
XMMU J013302.0$-$400628 & 3.023 & 2.10 $\pm$ 0.07 & 2.08 $\pm$ 0.10 &
1.90 $\pm$ 0.06 & 1.89 $\pm$ 0.03 \\ 
XMMU J013257.8$-$401030 & 2.180 & 1.67 $\pm$ 0.23 & 2.56 $\pm$ 0.25 & 1.91 $\pm$
0.25 & 1.94 $\pm$ 0.10 \\
XMMU J013314.7$-$401212 & 1.49 & 2.27 $\pm$ 0.58 & 2.42 $\pm$ 0.27 & 1.53 $\pm$
0.47 & 2.65 $\pm$ 0.17\\
XMMU J013205.4$-$400047 & 2.12 & 2.05 $\pm$ 0.16 & 2.55 $\pm$ 0.16 & 2.21 $\pm$
0.22 & 2.31 $\pm$ 0.08 \\
XMMU J013348.3$-$400233 & 2.11 & 3.48 $\pm$ 1.14 & 1.22 $\pm$ 1.03 & 1.47 $\pm$
1.14 & 2.33 $\pm$ 0.41  \\
XMMU J013233.2$-$395445 & 1.73 & 1.77 $\pm$ 0.38 & 2.20 $\pm$ 0.57 & 2.54 $\pm$
0.54 & 1.76 $\pm$ 0.19\\
XMMU J013333.8$-$395554 & 1.90 & 2.21 $\pm$ 0.32 & 2.47 $\pm$ 0.34 & 2.59 $\pm$
0.49 & 2.19 $\pm$ 0.15\\
XMMU J013408.6$-$400547 & 2.14 & 2.33 $\pm$ 0.25 & 2.64 $\pm$ 0.17 & 2.10 $\pm$
0.33 & 2.59 $\pm$ 0.10 \\
XMMU J013320.4$-$402021 & 2.25 & 1.37 $\pm$ 0.61 & 1.10 $\pm$ 1.57 & 1.49 $\pm$
0.60 & 1.33 $\pm$ 0.31 \\
&\\
& {\bf Mean} & {\bf 1.88~$\pm$~0.04} & {\bf 2.07~$\pm$~0.04} & {\bf
1.87~$\pm$~0.05} & {\bf 1.76~$\pm$~0.01} \\ 
\end{tabular}
\end{center}
\end{table*}

\section{High flux objects}
\label{sec:high}

Within this sample of AGN are 7 objects with 2--10~keV fluxes of
greater than 5~$\times~10^{-14}$~erg~cm$^{-2}$~s$^{-1}$. These have
spectra with sufficiently good signal-to-noise to allow individual 
analysis in greater detail.

The central object of the second field, QSO~0130$-$403
(XMMU~J013302.0-400628 in the tables), has the highest {\it
luminosity}, at 8.28~$\times~10^{45}$~erg~s$^{-1}$. The QSO was
discovered by Smith (1976) on Tololo survey plates and found to be
a very blue object, from the optical photometry published by Adam (1985).
It has previously been observed in the low-energy X-ray region by the 
{\it ROSAT} PSPC (Position Sensitive Proportional Counter), which 
gave a value of
$\Gamma$~=~1.69$^{+0.71}_{-0.60}$, when fixing N$_{H}$ to the Galactic
value (Bechtold \etal 1994).

\begin{figure}
\begin{center}
\includegraphics[clip,width=5.9cm,angle=-90]{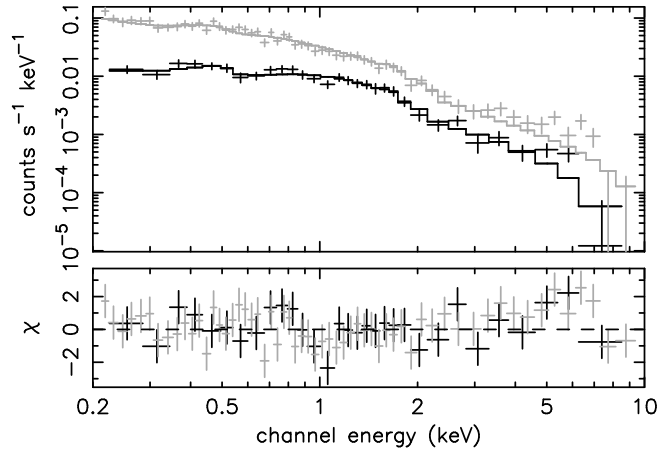}
\caption{The broad-band spectrum of QSO~0130$-$403 is best fitted by a
power law of $\Gamma$~$\sim$~2.09 and a broad Fe emission line at
6.6~keV (rest frame energy). The MOS spectrum is shown in black and the PN in grey.}
\label{Q_1}
\end{center}
\end{figure}

When fitting the 2--10~keV rest-frame band, 
the addition of an ionised Fe emission 
line, with the line width ($\sigma$) set to 0.1~keV, 
decreased $\chi^{2}$ by 5 for 2 degrees of
freedom, which corresponds to a marginal detection at $>$90~per~cent. 
(The width of the line could not be well constrained:
$\sigma$~=~0.34~$\pm$~0.29). 
Over the full 0.2--10~keV observed band (corresponding to 0.8--40.2~keV rest
frame), the best fit model consists simply of a power law
($\Gamma$~=~2.09~$\pm$~0.03) and the
emission line, detailed above, at 6.62~$\pm$~0.12~keV, the equivalent
width of the line being $\sim$200~eV in the QSO rest frame ($\sim$50~eV observed). 
There is no requirement for additional column density
or a soft excess. The spectrum is shown in Figure~\ref{Q_1}.

As mentioned in section~\ref{sec:specanal_speccurv}, there is
evidence for neutral reflection in this spectrum, with the 90~per~cent
confidence range for R being 1.11--3.85; i.e. consistent with
reflection from the full 2$\pi$ steradians of an accretion disc.

\begin{figure*}
\begin{center}
\includegraphics[clip,width=6.0cm,angle=-90]{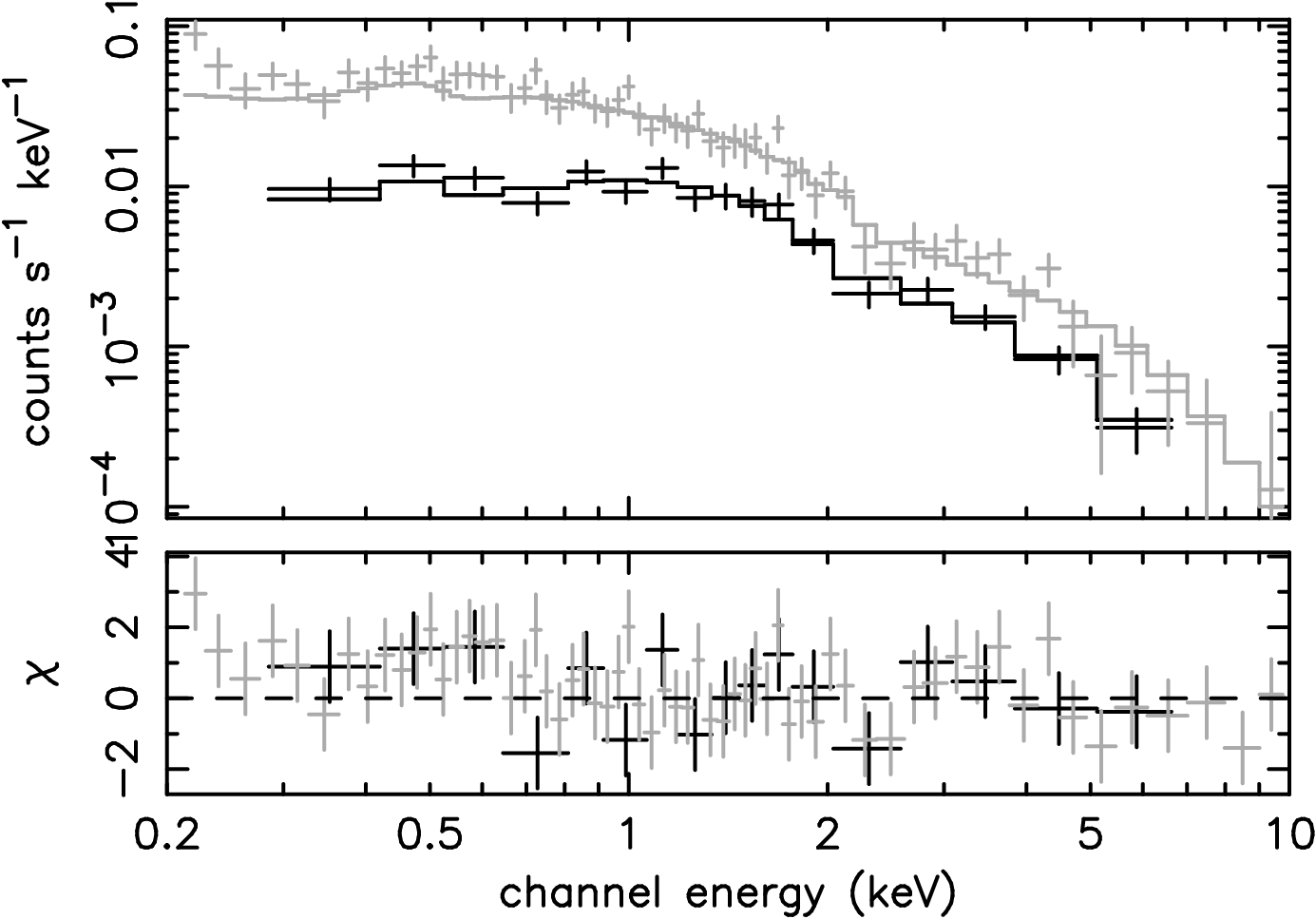}\hspace*{1cm}
\includegraphics[clip,width=6.0cm,angle=-90]{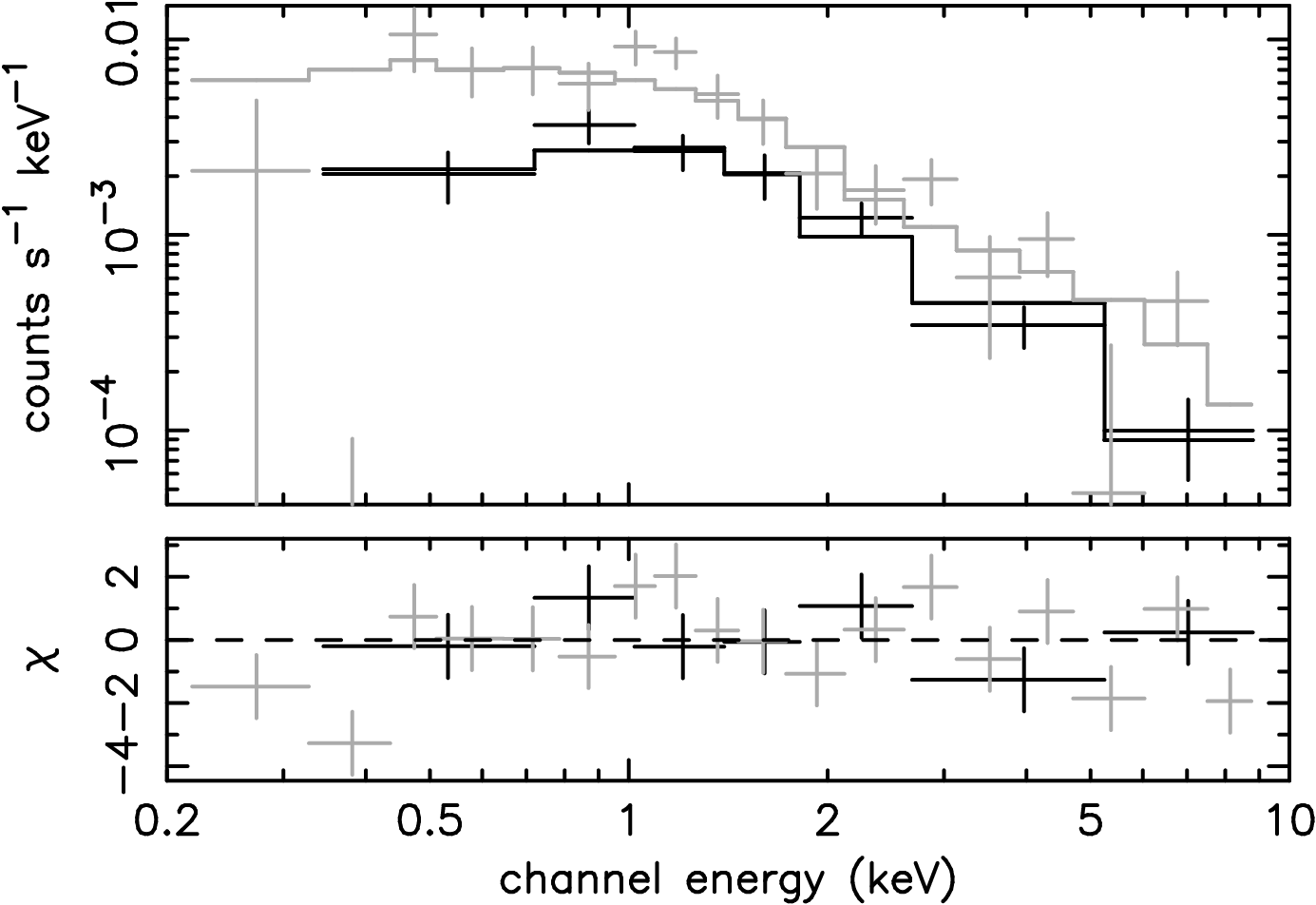}\vspace*{0.9cm}
\includegraphics[clip,width=6.0cm,angle=-90]{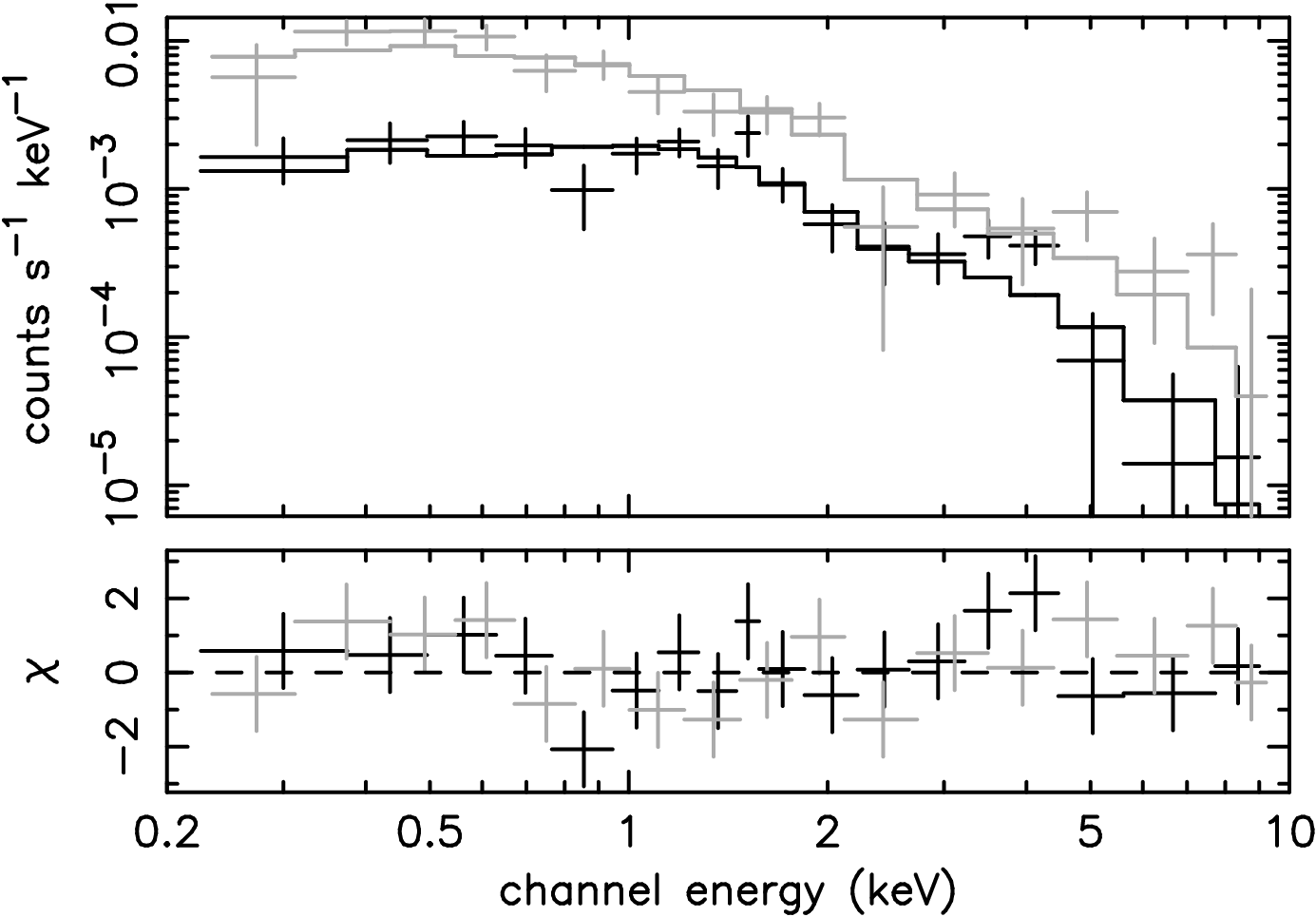}\hspace*{1cm}
\includegraphics[clip,width=6.0cm,angle=-90]{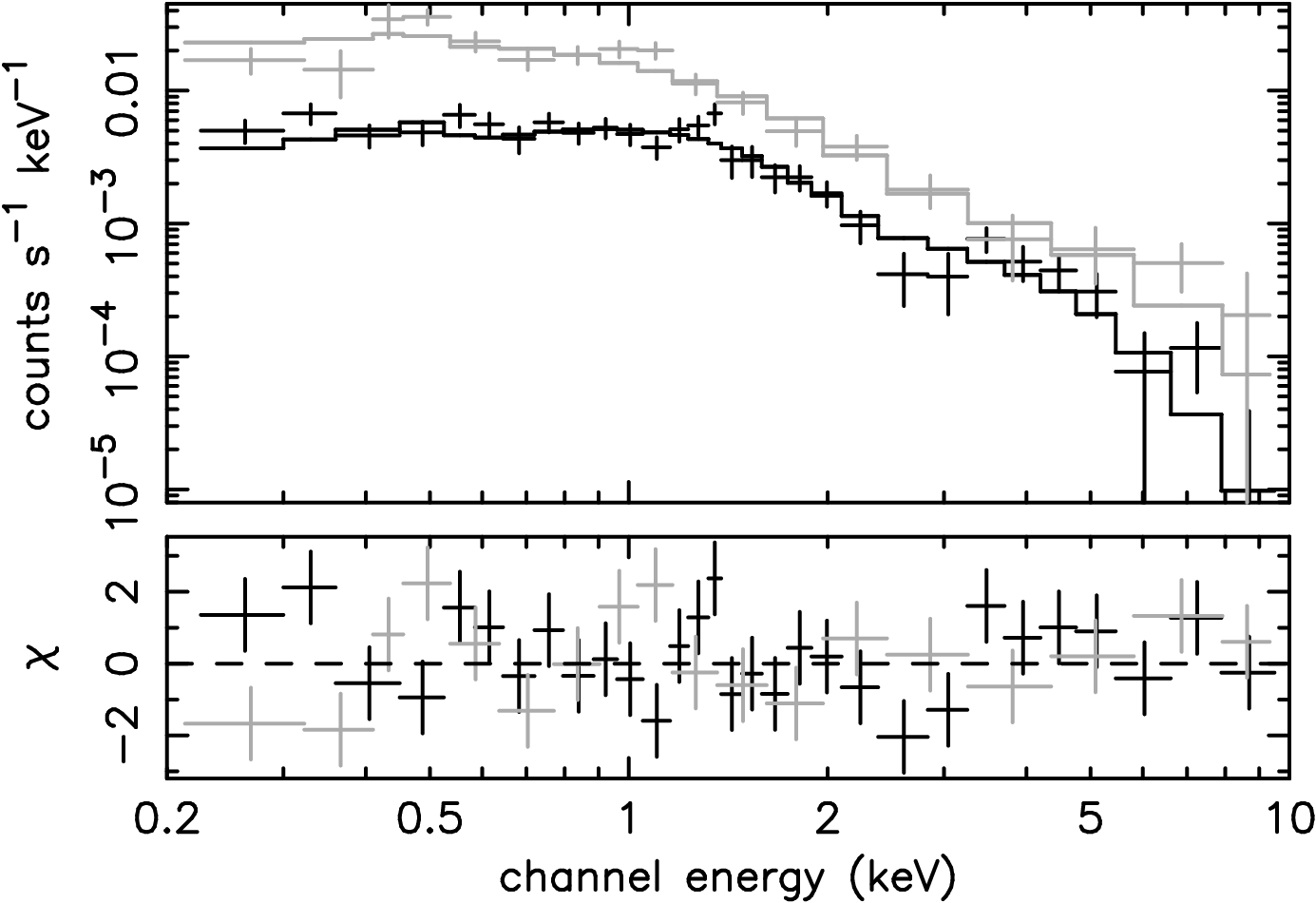}\vspace*{0.9cm}
\includegraphics[clip,width=6.0cm,angle=-90]{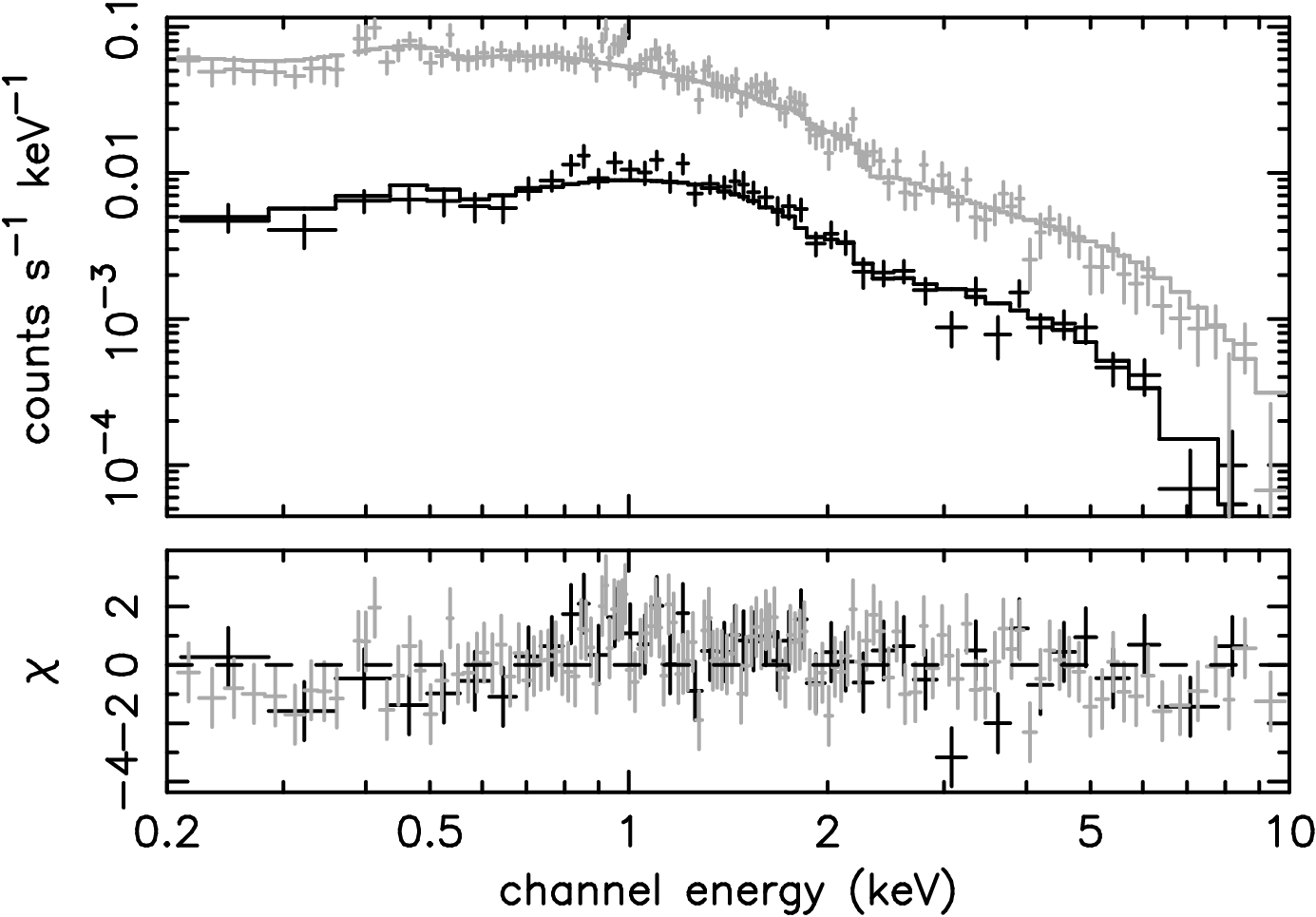}\hspace*{1cm}
\includegraphics[clip,width=6.0cm,angle=-90]{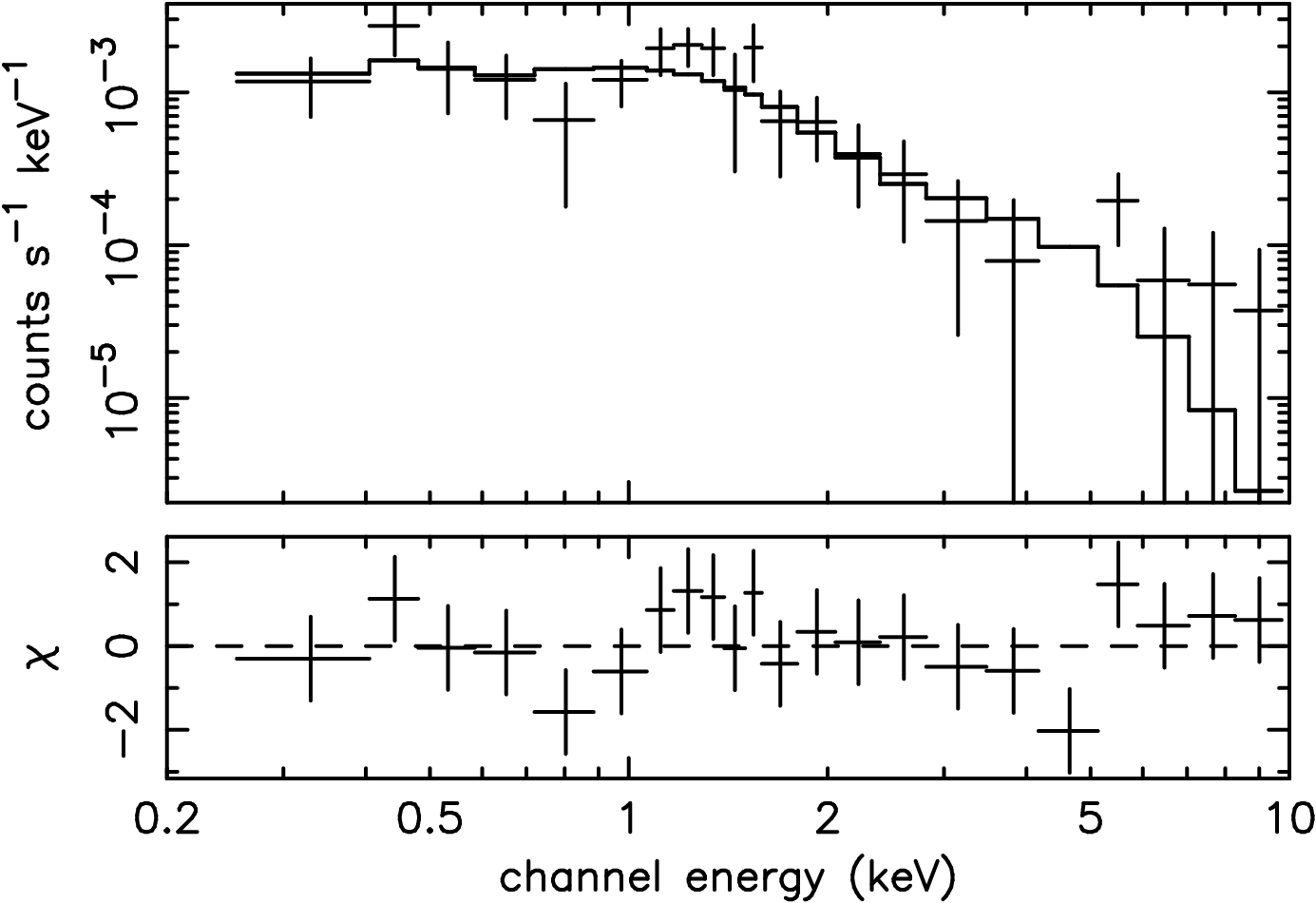}
\caption{Broad band spectra for those QSOs with a 2--10~keV flux of
greater than 5~$\times~10^{-14}$~erg~cm$^{-2}$~s$^{-1}$. From top to
bottom, the objects are XMMU~J122052.0+750529, XMMU~J122135.5+750914,
XMMU~J122206.4+752613 
in the left-hand column; XMMU~J122258.1+751934,
XMMU~J122351.0+752227 and XMMU~J013233.2-395445 on the right.}
\label{highfluxobjects}
\end{center}
\end{figure*}

As stated earlier, most of the spectra did not require the presence of
a strong soft excess. For each of the high flux objects, a blackbody component
was added into the model, with the temperature
fixed at 0.1~keV; the 90~per~cent upper limit was then calculated for the
luminosity over 0.2--10~keV; in general this was found to be $\leq$10 per cent of the 
strength of the power law. For a low-luminosity QSO, such as Mrk~205 
(Reeves \etal
2001), this is a typical value for the relative strength of the soft 
excess compared to the power law component; thus, the objects in
this sample are consistent with other low-luminosity
QSOs, where better signal-to-noise in the spectra allow more detailed
analysis of the soft excesses.

All the high-flux spectra (Figure~\ref{highfluxobjects}) 
were systematically fitted to search for the presence
of iron lines and reflection. 
With the exception of QSO~0130$-$403, neither iron emission nor
the Compton Reflection Hump could be constrained in any of the 
individual spectra, with 90 per cent upper limits on the equivalent
widths of the iron lines
lying between 145--650~eV.

It should be noted that two of these bright objects are those
classified as Narrow Line AGN by XB02 -- XMMU~J122206.4+752613 and
XMMU~J122258.1+751934. These are the only two high-flux objects which 
have measureable
excess absorption; the values, however 
($\leq$~few~$\times~10^{21}$~cm$^{-2}$), are not of a large enough
magnitude completely to 
obscure broad-line regions and, hence, to explain the narrow-line classification. 

Since Mrk~205 has been presented in an earlier paper, its spectrum has
not been re-fitted. However, from Reeves \etal (2001), we find that
the spectral slope for Mrk~205
is 1.80~$\pm$~0.04, for the energy band 2.5--10~keV. This includes two
Gaussian components: a narrow ($\sigma$~=~10~eV) line at
E~=~6.39~$\pm$~0.02~keV and a broader ($\sigma$~=~250$^{+190}_{-130}$~eV)
one at E~=~6.74~$\pm$~0.12~keV. The broad-band spectrum shows a slight
soft excess, and is best fitted by $\Gamma$~=~1.86~$\pm$~0.02 and a
blackbody component with kT~=~120~$\pm$~8~eV, together with the two
emission lines. It is clear that these values for the slopes are
consistent with those for the objects presented in this paper.

\section{Discussion}
\label{sec:disc}
\subsection{Absorbing Column Density}
\label{sec:disc_nh}

Our fits to the QSO spectra implied that the RQQs 
observed have no additional absorption above that due to 
our own Galaxy. 
The flatter spectral slopes for some objects could not be explained by
obscuration in the AGN; this agrees with the {\it XMM} observation of
the Lockman Hole (Hasinger \etal 2001; Mainieri \etal 2002), in that
our sources are type-1 AGN, while their intrinsically absorbed objects are mainly type-2. One possibility is that we are not
reaching a low enough ($<10^{-14}$~erg~cm$^{-2}$~s$^{-1}$) 
flux level to be measuring spectra of these faint, obscured sources, 
within these two fields. 

However, the lack of any absorption above the 
Galactic value agrees with previous studies of
radio-quiet QSOs, e.g., with {\it Einstein} (Canizares \& White
1989; objects with 0.1$<$z$<$3.5), {\it ROSAT} (Yuan \etal
1998; z$>$2) and {\it Ginga} (Lawson \& Turner 1997; z$<$1.4). Indeed
the radio-quiet QSOs in this sample appear to remain unabsorbed up to 
redshifts of z~$\sim$~3.
 
Radio-loud objects, on the other hand, do tend to show excess
absorption (e.g., Elvis \etal 1994; Brinkmann, Yuan \& Siebert 1997;
RT00). The data from RT00 show
an average column density of 1.29~$\times~10^{22}$~cm$^{-2}$ for radio-loud 
quasars. For the sample of RQQs in this paper, a mean of
2.96~$\times~10^{21}$~cm$^{-2}$ is found for the {\it upper limits}, 
which is noticeably smaller. 

It is known that the low-energy response of the {\it ASCA} SIS has
been degrading over time (see Yaqoob \etal 2000). 
The effect of this is an underestimate of
the soft X-ray flux, which can show itself as an increase in
N$_{H}$. In order to check that the differences found between the
RLQ and RQQ distributions are not instrument-linked, the
measured column densities for two objects which have
been observed by both {\it XMM} and {\it ASCA} were compared ({\it
ASCA} values were taken from RT00; {\it XMM} values calculated for own
data for PKS~0558$-$504 and taken from Reeves \etal 2001 for Mrk~205). It was
found that, within the errors, the instruments agreed; it can,
therefore, be assumed that the absorption measured for the RT00
radio-loud sample is not due simply to the {\it ASCA} calibration.

\begin{figure}
\begin{center}
\includegraphics[clip,width=6.0cm,angle=-90]{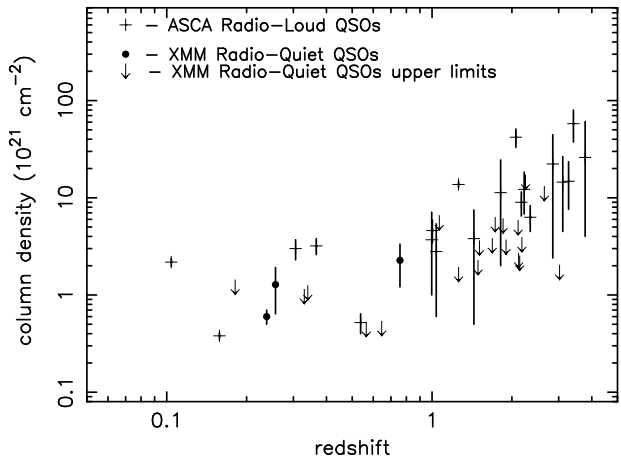}
\caption{The joint sample of radio-quiet and radio-loud QSOs, from
{\it XMM} and {\it ASCA} (taken from Reeves \& Turner 2000) respectively.}
\label{RQ+L}
\end{center}
\end{figure}

Statistical tests were applied to both the radio-quiet objects
from this paper and the radio-loud ones from RT00, to determine 
whether there was evidence for a difference in
population. Fig.~\ref{RQ+L} shows the column densities against redshift for
this combined sample, with the {\it ASCA}-detected RLQs marked by
crosses and the {\it
XMM} RQQs by circles/arrows.

Due to the overwhelming presence of censored data 
(that is, upper limit measurements), the {\it ASURV}
(Astronomy Survival Analysis) software was used (see Feigelson \&
Nelson 1985). Employing each of the possible two-sample tests, we
found that the probability that the RQQs
and RLQs are from the same population was very low: $<$0.01~per~cent.
The Kolmogorov--Smirnov test was also performed on the data, leading to
a probability of $\sim$3.4~per~cent for the existence of one population only. 
Other statistical tests were also tried, each method confirming 
the same result: 
the radio-quiet QSOs in our sample were found to belong to a different population from the radio-loud objects in
RT00 (probability of $>$99.99~per~cent in general).

Summarising, we find that, although RLQs tend to show excess
absorption at high redshifts, RQQs do not, at least up to z~$\sim$~3. 
This 
implies that there is an intrinsic difference in the local environment
of the two types of 
QSO at high redshifts. It must be noted that the situation for radio-quiet QSOs at higher
redshifts (z~$>$~5) is unclear; deep {\it XMM} observations of
individual, bright high-redshift QSOs are required to explore this 
possibility.

\subsection{Spectral Evolution}
\label{sec:disc_specevol}

The QSOs in this sample all show the same range of continuum
photon indices, regardless of luminosity or redshift, over the 
energy range 2--10~keV. Comparing the slopes over the full 0.2--10~keV
range gave the same results.

Canizares \& White (1989) also find no evidence for a $\Gamma$-z dependence
from the {\it Einstein} data, up to a maximum energy of 4~keV. Likewise,
Yuan \etal (1998) found that the
spectral slopes of distant (z~$>$~2.5) RQQs were consistent (within the errors)
with those of nearby objects; {\it ROSAT} covered the energy range
0.1--2.4~keV. The {\it ASCA} data in RT00 also
showed no significant correlation between $\Gamma$ and redshift, over
a similar energy range to that of {\it XMM}.
The independence of spectral slope and redshift is also observed in
the higher energy band (E$\geq$2~keV) of
radio-loud objects 
[e.g., comparison between $\Gamma$ from {\it EXOSAT} data (Lawson
\etal 1992) and {\it Ginga} (Williams \etal 1992)].
There also appears to be no evolution with redshift of the extreme-ultraviolet
spectral slope, for either RL or RQQs (Telfer \etal 2002).

The link between the different regions of the spectral energy
distributions of AGN is, as yet, uncertain, with no complete theory having
been generally accepted. The observed lack of correlation between slope and
redshift may suggest, however, that the luminosity of these QSOs is
simply scaling as the black hole mass -- i.e. $\dot{M}$ remains
constant with respect to the Eddington value, as the overall spectral
energy distribution remains unchanged. Different values of the
accretion rate could then be responsible for some of the dispersion of
the spectral slopes that is observed (e.g., Liu, Mineshige \& Shibata 2002). 
Alternatively, the spread of $\Gamma$ may be due
to different disc-corona geometries 
(Czerny \& Elvis 1987; Fabian \etal 2002). It should be noted that this
dispersion in slope is also observed in the low-luminosity Seyferts,
as well as QSOs: Narrow Line Seyfert 1 galaxies tend to have softer spectra
than the corresponding broader line objects; this may be explained 
if these objects are accreting at a greater fraction of the 
Eddington rate, when compared to the broad line AGN (Pounds, Done \&
Osborne 1995).

\subsection{Spectral Curvature}
\label{sec:disc_speccurv}

For the signal-to-noise of the data presented
here, it is not possible to make a conclusive
statement on the presence or absence of reflection in the 
{\it individual} QSOs in this sample. 
However, the spectral hardening (where $\Delta\Gamma=0.2$) 
that appears to exist for the sample as a whole above 3~keV, is consistent with reflection from 2$\pi$ steradians of 
optically thick matter, located out of the line of sight.

Past studies (e.g., with {\it ASCA}, RT00) have noted that the strength
of the neutral Compton reflection hump, and the associated iron K$\alpha$ 
line, is considerably lower in the higher-redshift QSOs, than 
that expected in Seyfert 1 galaxies at low redshift (e.g., Nandra 
\& Pounds 1994). However, the QSO spectra previously obtained above z~=~2 
by {\it ASCA} were dominated by core-dominated radio-loud quasars, where 
a contribution from the relativistic jet could weaken any apparent 
reflection features. 
This does not appear to be the case for the {\it radio-quiet} AGN 
found in this sample; the average co-added spectra 
of these AGN above z~=~2 appears to indicate the presence of a reflection component 
with R~$\sim$~1, the first time this has been observed for high-redshift 
QSOs. Such a reflection component could naturally arise from X-rays 
scattering off optically thick material, such as the inner accretion disc, 
subtending $\sim2\pi$ steradians to the X-ray source. 
Another possible contribution is by reflection 
off a distant molecular torus, predicted from AGN unification schemes.  
We note here that such a component may account for the `narrow' 
iron K$\alpha$ lines observed by {\it XMM-Newton} and {\it Chandra} 
in bright, nearby AGN (e.g., Kaspi \etal 2001; O'Brien \etal 2001a). 
 
Finally we note the possibility that the reflection may 
occur from an ionised disc (e.g., Reeves \etal 1997; Eracleous \&
Halpern 1998; Grandi \etal 2001; Ballantyne, Ross \& Fabian 2002).
However, the data in this paper show only one significant iron line, 
located near 6.6~keV; this tentatively 
suggests that the reprocessing material
might be more strongly ionised in some of the higher luminosity AGN, 
as is observed for instance in the {\it XMM-Newton} 
spectrum of Mrk~ 205 (Reeves \etal 2001).  Such an ionised disc could
also explain some of the observed spectral curvature (e.g., O'Brien
\etal 2001b; Pounds \etal 2001). 

\section{Conclusions}

Analysing 23 radio-quiet QSOs, with a mean redshift of 1.40 and
2--10~keV luminosity of 9.66~$\times~10^{44}$~erg~s$^{-1}$, 
we have found the following:

\begin{itemize}
\item The 2--10~keV spectral slopes measured for this sample
show a significant scatter. However there is 
no obvious trend with redshift, so
it can be concluded that there is no evidence for X-ray spectral evolution
of QSOs with redshift or luminosity
\item The mean value for $\alpha_{ox}$ is measured to be
$-$1.66~$\pm$~0.04. Within the sample there is no correlation with
redshift, although comparison with both lower and higher redshift
samples indicates a slight steepening as z increases.  
\item The dispersion in photon index is intrinsic; that is, it is not related
to obscuration, as is thought to be the case for the faint sources
constituting X-ray background.
\item No excess absorption in the rest frame of the radio-quiet
QSOs is found. This is in contrast to the additional N$_H$ measured
by past studies in high-z radio-loud quasars and suggests an intrinsic
difference could exist in the local environments of the two classes
of QSO.
\item Considering the sample as a whole, the spectral slope appears to
flatten at higher energies (above 3 keV).  
This indicates the presence of a
reflection component from material 
subtending a solid angle of 2$\pi$ steradians to the X-ray source. 
This matter could arise from either 
the inner accretion disc or a distant ($>$1~pc) molecular torus.

\end{itemize}

\section{ACKNOWLEDGMENTS}
The work in this paper is based on observations with {\it
XMM-Newton}, an ESA
science mission, with instruments and contributions directly funded by
ESA and NASA. The authors would like to thank the EPIC Consortium for all their work during the calibration phase, 
and the SOC and SSC teams for making the observation and analysis
possible. Thanks also go to Xavier Barcons and the AXIS team for the
data on the QSOs in the field of Mrk~205, together with the
anonymous referee, for very helpful comments.
This research has made use of the NASA/IPAC Extragalactic
Database (NED), which is operated by the Jet Propulsion Laboratory,
California Institute of Technology, under contract with the National
Aeronautics and Space Administation. For the survival statistics, the {\it
ASURV} v1.1 (Astronomy Survival Analysis) software was utilised
(Feigelson \& Nelson 1985). 
Support from a PPARC studentship and the Leverhulme Trust is
acknowledged by KLP and JNR respectively.


\begin{thebibliography}{}

\bibitem[Adam 1985]{ad85}Adam G., 1985, A\&AS, 61, 225
\bibitem[Antonucci 1993]{an93}Antonucci R., 1993,
ARA\&A, 31, 473
\bibitem[Ballantyne 2002]{bal02}Ballantyne D.R., Ross R.R., Fabian
A.C., 2002, MNRAS, 332, L45
\bibitem[Barcons 2001]{ba01}Barcons X. \etal, 2002, A\&A, 382, 522 (XB02)
\bibitem[Bechtold 1994]{be94}Bechtold J. \etal, 1994, AJ, 108, 759 
\bibitem[Brinkmann 1997]{br97}Brinkmann W., Yuan W., Siebert J., 1997,
A\&A, 319, 413
\bibitem[Canizares 1989]{ca89}Canizares C.R., White J.L., 1989, ApJ,
339, 27
\bibitem[Czerny 1987]{cz87}Czerny B., Elvis M., 1987, ApJ, 321, 305
\bibitem[Dickey 1990]{di90}Dickey J.M., Lockman F.J., 1990, ARA\&A,
28, 251
\bibitem[Elvis 1994]{el94}Elvis M., Fiore F., Wilkes B., McDowell J.,
Bechtold J., 1994, ApJ, 422, 60
\bibitem[Eracleous 1998]{er98}Eracleous M., Halpern J.P., 1998, ApJ,
505, 577
\bibitem[Fabian 2002]{fa02}Fabian A.C., Ballantyne D.R., Merloni A., Vaughan
S., Iwasawa K., Boller Th., 2002, MNRAS, 331, L35
\bibitem[Falomo 2001]{fa01}Falomo R., Kotilainen J., Treves A., 2001,
ApJ, 547, 124
\bibitem[Feigelson 1985]{fe85}Feigelson E.D., Nelson P.I., 1985, ApJ,
293, 192
\bibitem[Grandi 2001]{gr01}Grandi P., Maraschi L., Urry C.M., Matt G.,
2001, ApJ, 556, 35
\bibitem[Hasinger 2001]{ha01}Hasinger G. \etal, 2001, A\&A, 365, L45
\bibitem[Kaspi 2001]{ka01}Kaspi S. \etal, 2001, ApJ, 554, 216 
\bibitem[Kellerman 1989]{ke89}Kellerman K.I., Sramek R., Schmidt M.,
Shaffer D.B., Green R., 1989, AJ, 98, 1995
\bibitem[Kukula 1998]{ku98}Kukula M.J., Dunlop J.S., Hughes D.H.,
Rawlings S., 1998, MNRAS, 297, 366
\bibitem[Lamer 1997]{la97}Lamer G., Brunner H., Staubert R., 1997,
A\&A, 327, 467
\bibitem[Lawson 1992]{la92}Lawson A.J., Turner M.J.L., Williams O.R.,
Stewart G.C., Saxton R.D., 1992, MNRAS, 259, 743
\bibitem[Lawson 1997]{la97}Lawson A.J., Turner M.J.L., 1997, MNRAS,
288, 920
\bibitem[Liu 2002]{li02}Liu B.F., Mineshige S., Shibata K., 2002, ApJ,
572, L173
\bibitem[Maccacaro 1988]{ma98}Maccacaro T., Gioia I.M., Wolter A.,
Zamorani G., Stocke J.T., 1988, ApJ, 326, 680
\bibitem[Magdziarz 1995]{ma95}Magdziarz P.,  Zdziarski A.A., 1995,
MNRAS, 273, 837
\bibitem[Mainieri 2002]{ma02}Mainieri V., Bergeron J., Hasinger G.,
Lehmann I., Rosati P., Schmidt M., Szokoly G., Della Ceca R., 2002,
A\&A, in press (astro-ph/0207166)  
\bibitem[Miller 1990]{mi90}Miller L., Peacock J.A., Mead A.R.G., 1990,
MNRAS, 244, 207
\bibitem[Nandra 1994]{nan94}Nandra K., Pounds K.A., 1994, MNRAS, 268, 405
\bibitem[O'Brien 2001a]{ob01a}O'Brien P.T., Page K., Reeves J.N., Pounds
K., Turner M.J.L., Puchnarewicz E.M., 2001a, MNRAS, 327, L37
\bibitem[O'Brien 2001b]{ob01b}O'Brien P.T., Reeves J.N., Turner
M.J.L., Pounds K.A., Page M., Gliozzi M., Brinkmann W., Stephen J.B.,
Dadina M., 2001b, A\&A, 365, L122
\bibitem[Pounds 1990]{po90}Pounds K.A., Nandra K., Stewart G.C.,
George I.M., Fabian A.C., 1990, Nat, 344, 132
\bibitem[Pounds 1995]{po95}Pounds K.A., Done C., Osborne J., 1995,
MNRAS, 277, L5
\bibitem[Pounds 2001]{po01}Pounds K., Reeves J., O'Brien P., Page K.,
Turner M., Nayakshin S., 2001, ApJ, 559, 181
\bibitem[Pounds 2002]{po02}Pounds K.A., Reeves J.N., 2002, in  
New Visions of the X-ray Universe in the {\it XMM-Newton} and {\it
Chandra} era (astro-ph/0201436)
\bibitem[Reeves 1997]{re97}Reeves J.N., Turner M.J.L., Ohashi T., Kii
T., 1997, MNRAS, 292, 468
\bibitem[Reeves 2000]{re00}Reeves J.N., Turner M.J.L., 2000, MNRAS,
316, 234 (RT00)
\bibitem[Reeves 2001]{re01}Reeves J.N., Turner M.J.L., Pounds K.A.,
O'Brien P.T., Boller Th., Ferrando P., Kendziorra E., Vercellone S.,
2001, A\&A, 365, 134
\bibitem[Smith 1976]{sm76}Smith M.G., 1976, ApJ Letters, 206, L125
\bibitem[Str\"{u}der 2001]{st01}Str\"{u}der L. \etal, 2001, A\&A, 365,
L18
\bibitem[Telfer 2002]{te02}Telfer R.C., Zheng W., Kriss G.A., Davidsen
A.F., 2002, ApJ, 565, 773
\bibitem[Turner 2001]{tu01}Turner M.J.L. \etal, 2001, A\&A, 365, L27 
\bibitem[Veron 1990]{ve90}V\'{e}ron-Cetty M.-P., Woltjer L., 1990,
A\&A, 236, 69
\bibitem[Veron 2001]{ve01}V\'{e}ron-Cetty M.-P., V\'{e}ron P., 2001,
A\&A, 374, 92 
\bibitem[Vignali 2001]{vi01}Vignali C., Brandt W.N., Fan X., Gunn
J.E., Kaspi S., Schneider D.P., Strauss M.A., 2001, AJ, 122, 2143
\bibitem[Wilkes 1987]{wi87} Wilkes B.J., Elvis M., 1987, ApJ 323, 243
\bibitem[Williams 1992]{wi92}Williams O.R. \etal, 1992, ApJ, 389, 157
\bibitem[Worrall 1987]{wo87}Worrall D.M., Giommi P., Tananbaum H., Zamorani G., 1987,
ApJ, 313, 596
\bibitem[Yaqoob 2000]{ya00}Yaqoob T. \etal, 2000, {\it ASCA} Guest
Observer Facility Calibration Memo, ASCA-CAL-00-06-01 (http://asca.gsfc.nasa.gov/docs/asca/calibration/nhparam.html)
\bibitem[Yuan 1998]{yu98}Yuan W., Brinkman W., Siebert J., Voges W.,
1998, A\&A, 330, 108
\bibitem[Zamorani 1981]{za81}Zamorani G. \etal, 1981, ApJ, 245, 357

\end{thebibliography}
\end{document}